\documentclass[letterpaper, 10pt, conference]{ieeeconf}
\IEEEoverridecommandlockouts 
\usepackage{amsmath}
\DeclareMathOperator*{\argmax}{arg\;max}
\DeclareMathOperator{\subjectto}{subject\;to}
\DeclareMathOperator*{\re}{Re}
\DeclareMathOperator*{\im}{Im}
\DeclareMathOperator*{\sign}{sign}
\usepackage{graphicx}
\usepackage{url}
\usepackage{algorithm}
\usepackage[caption=false,font=footnotesize]{subfig}
\usepackage{algorithmic}

\begin{document}
\title{Power Injection Attacks in Smart Distribution Grids with Photovoltaics}


\author{Martin Lindstr\"{o}m, Hampei Sasahara, Xingkang He, Henrik Sandberg, and Karl Henrik Johansson%
    \thanks{The work was supported by Knut \& Alice Wallenberg foundation and the Swedish Research Council.}
    \thanks{The authors are all with the Division of Decision and Control Systems, KTH Royal Institute of Technology, Stockholm, Sweden. E-mails: {\tt \{mali5, hampei, xingkang, hsan, kallej\}@kth.se}.}
}
    
\maketitle

\begin{abstract}
In order to protect smart distribution grids from intrusions, it is important to understand possible risks and impacts of attacks. We study the worst-case attack strategy of a power injection attack against the physical layer of a smart distribution grid with a high penetration of photovoltaic resources. We derive both the worst attack signal and worst attack location: The worst attack signal is a step function which switches its sign at the final stage, and the worst attack location is the node with the largest impedance to the grid substation. Numerical examples on a European benchmark model verify the developed results. Finally, both theoretical and numerical results are used to discuss feasible defense strategies against power injection attacks.
\end{abstract}

\section{Introduction}
The integration of communication and computational capabilities of a cyber system, with a physical or engineered system, results in a cyber-physical system (CPS). A typical CPS would use the cyber layer and feedback loops to control the physical layer \cite{Lee:CPS_design_challenges}. CPS security has been a natural and crucial consideration in recent years \cite{Giraldo:surveyofsurveys, Humayed:CPS_survey, Sandberg:CPS_networked_control, Dibaji:Control_perspective_CPS}. The smart grid (SG), a traditional 20-th century power grid augmented with sensors, actuators and cyber components, is a typical class of CPS. 

SG security is considered to be one of the most important topics of CPS security research \cite{Giraldo:surveyofsurveys, Humayed:CPS_survey}, and has been considered in e.g. \cite{sridhar:cps_powergrid, li:security_distributed_SG, he:cps_defense_SG, musleh:detection_FDIA_SG,  konstantinou:hardware_SG}. The introduction of a cyber layer into the power grid aids the grid operator in regulating the power grid. However, cyber components also create vulnerabilities for an attacker to exploit. One well-known power grid security breach is the cyber attack on the Ukranian power grid in 2015 that resulted in approximately $225\:000$ consumers losing power \cite{ukraine_attack}. Though the attack caused no operational impact on critical infrastructure, the attack highlighted the importance of power grid security.




Power grids generally, and SGs specifically, have two different functions: transmission and distribution. Due to the need for clean energy resources, a common topic in SG research is smart distribution grids with a high penetration of photovoltaic (PV) resources. PV resources are often equipped with direct current/alternating current (DC/AC) inverters with variable reactive power generation, through which, the grid operator can realize a control law. Voltage control in a smart distribution grid with inverter-equipped PV resources has been considered in \cite{Andren:stability_inverter-based, Turitsyn:reactive_power_flow_control, Chong:initial_droop}. One could realize many different control laws through the inverters, and it is worth noting that \cite{Andren:stability_inverter-based, Turitsyn:reactive_power_flow_control, Chong:initial_droop} use different control laws.


In this paper, we study a power injection attack into a branch of a smart distribution grid with a high penetration of PV resources, which can be regarded as a deception attack \cite{Dibaji:Control_perspective_CPS}, targeted against the physical layer of the smart distribution grid. In this scenario, the attacker has the capability to inject active and/or reactive power into the grid, but the attack is constrained by a maximum attack magnitude and a maximum attack length. The attacker's objective is to achieve maximal voltage deviation by injecting power into the grid in order to force the grid operator to shut down some part of the grid. On the other hand, the grid operator can use the inverters of the PV resources to regulate the produced reactive power, which in turn can be used to regulate voltage levels in the grid in order to counteract the attack. Our objective is to investigate the impact caused by the worst attack scenario, and from that, draw conclusions about viable defense strategies. Characterizing the worst case attack scenario is an important step of risk analysis and mitigation. Due to the importance of security in smart distribution grids, problems such as attack detection, secure grid design, and grid reconfiguration need to be considered. A foundation for such considerations is an understanding of the worst case attack scenario of a deception attack against the physical layer of a smart distribution grid. 

Our main contributions in this paper are twofold: First, we derive an explicit expression of the worst case attack, and we specify the most vulnerable point of an arbitrary smart distribution grid with radial (i.e. tree-structured) topology. The worst case attack signal for a finite-time attack is on the form of a step function whose sign switches at the final attack stage. The most vulnerable node in a radial grid is the node with the largest impedance to the substation. Second, the obtained results help motivate effective defense strategies. For a given voltage deviation, it is difficult in general to determine if the deviation is caused by the worst case attack scenario, or by normal grid operation. Hence, we focus our discussion on defense strategies on grid design and grid reconfiguration algorithms. 

There are a number of works studying attacks against the cyber layer in the context of SGs. Recent papers have proposed many analysis approaches, for example deep learning \cite{niu:deep_learning}, model-based diagnosis \cite{hampei:ifac, jiang:kalman_filter}, and reinforcement learning \cite{kurt:reinforcement_learning}. We provide an analysis of attacks from a physical layer perspective, which coupled with existing approaches yields additional insights into attack detection, grid design, and grid reconfiguration.

The paper is organized as follows. In Section~\ref{sec:pf}, a model of the grid is derived and a formal problem statement is presented. In Section~\ref{sec:attack}, the grid model is used to analyze the worst-case attack strategy. In Section~\ref{sec:simulations}, our results are verified with numerical examples, and the defender's perspective is briefly considered. We conclude with Section~\ref{sec:conclusion}.

\textbf{Notation}. The $N$-dimensional identity matrix is denoted by $I_N$, and define the $i$-th basis vector $e_i$ as the $i$-th column of $I_N$. A diagonal matrix with diagonal elements $\{d_1,\dots,d_k\}$ is denoted by $\text{diag}(d_1,\dots,d_k)$. The superscript ``${\sf T}$'' denotes the transpose of a vector or a matrix. $[M]_{i,j}$ denotes the element on row $i$ and column $j$ of the matrix $M$. Denote $M\succ 0$ ($M\prec 0$) if $M$ is a symmetric positive definite (negative definite) matrix. The function $\sign(\cdot)$ denotes the sign function. The imaginary unit is denoted by $j$. The real and imaginary parts of a complex number $z$ are denoted by $\re(z)$ and $\im(z)$, respectively.


\section{Model Description and Problem Formulation} \label{sec:pf}
In this section, we present a model of the grid, present the considered attack scenario, and finally present a mathematical problem formulation.


\subsection{Model Description} \label{subsec:model}
\begin{figure}
    \centering
    \includegraphics[width=\linewidth]{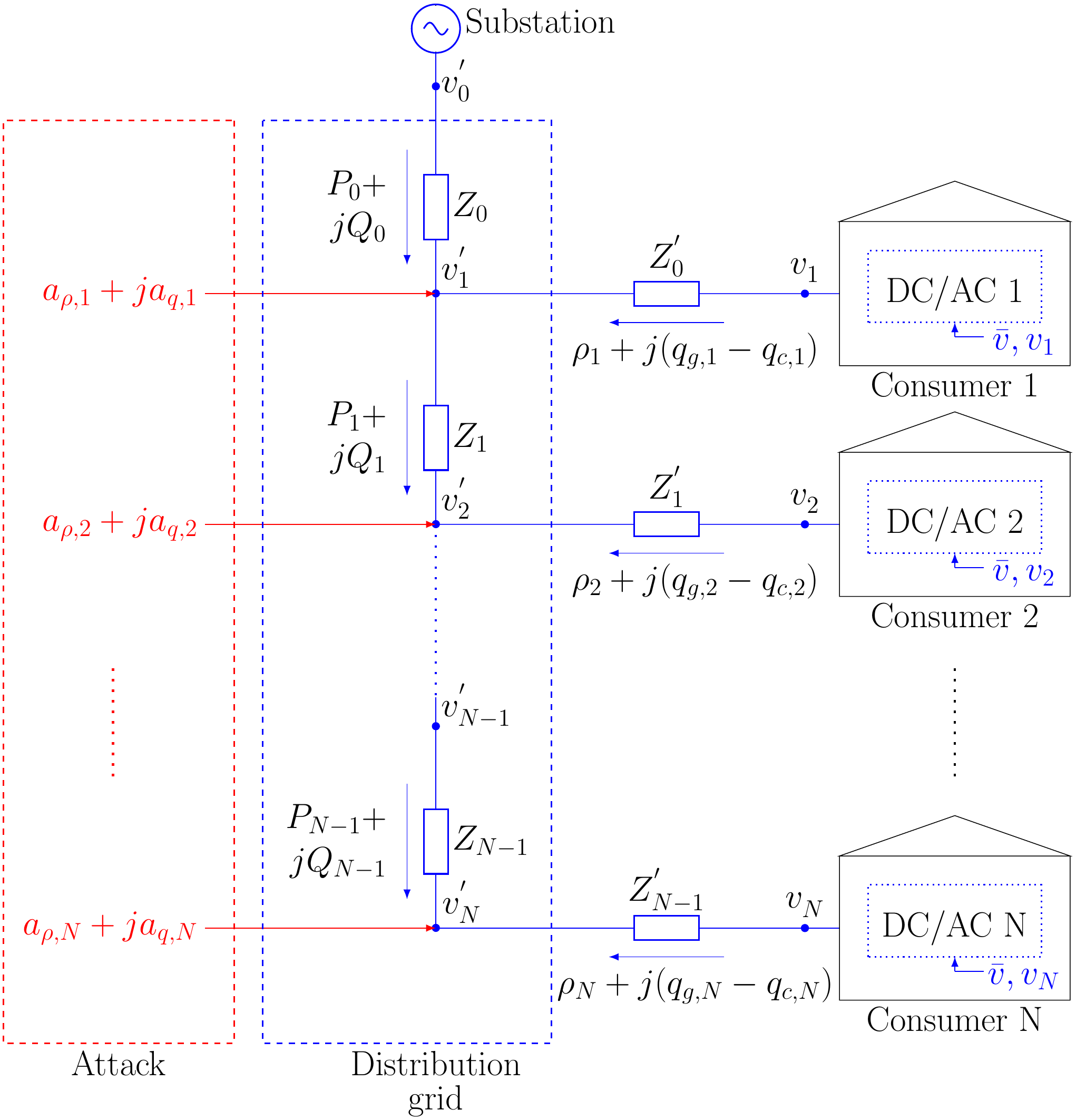}
    \caption{A line grid with $N$ consumers where each consumer has a PV resource. The grid could be attacked at the connection point of any consumer.}
    \label{fig:grid}
\end{figure}

Consider the smart distribution grid illustrated in Fig.~\ref{fig:grid}; a substation feeds consumers $i\in \{1,2,\dots,N\}$ who all have PV resources. This grid topology is a special case of a radial grid: a tree-structured grid without branching points. The grid in Fig.~\ref{fig:grid} is referred to as a \emph{line grid}. Consumer $i$ is connected to the main feeder through a line with impedance $Z'_{i-1}$, and the main feeder has impedance $Z_i$ between each consumer $i-1$ and consumer $i$'s connection points. It is assumed that $R_i:= \re(Z_i)$, $R'_i := \re (Z'_i)$, $X_i := \im(Z_i)$, and $X'_i := \im(Z'_i)$ are positive for all impedances. Active and reactive powers flowing in the main feeder are represented by $P_i$ and $Q_i,$ respectively. All consumers consume constant reactive power $q_{c,i}$ and use a PV resource with an associated inverter. It is assumed that the inverter generates constant active power $\rho_i$ and variable reactive power $q_{g,i}$. Define $\rho_i$ and $q_{g,i}$ to have positive sign when flowing out from the consumer, and define $q_{c,i}$ to have positive sign when flowing towards the consumer. The attacker could inject active power $a_{\rho,i}$ and reactive power $a_{q,i}$ at any of the $N$ consumer connection points. The attack scenario considered in this paper is described in detail in Section~\ref{subsec:attack_scenario}.

The linearized \textit{DistFlow} model in \cite{DistFlow} is used for describing power flow and voltage drop. The model is based on two fundamental concepts: one for nodes, and one for edges in the grid. For nodes, the net apparent power in each node must be $0$, and for edges, the voltage drop across an impedance is proportional to the power that flows through it. This yields the equations  
\begin{equation}\label{eq:base_eqns}
    \left\{
    \begin{aligned}
        P_{i+1} &= P_i + \rho_{i+1}+a_{p,i+1},\\
        Q_{i+1} &= Q_i + q_{g,i+1} - q_{c,i+1} + a_{q,i+1},\\
        {v'}_{i+1}^2 &= {v'}_{i}^2 - 2\beta_i (P_i,Q_i),\\
        v_i^2 &= {v'}_i^2 + 2\beta'_{i-1} (\rho_{i+1}, q_{g,i+1} - q_{c,i+1}),
    \end{aligned}\right.
\end{equation}
where $\beta_i (r,s) := R_i r + X_i s$ for $i\in \{0,1,\dots,N-1\}$ and $\beta'_{i} (r, s) := R'_{i} r + X'_i s$ for $i\in \{1,2,\dots,N\}$. Furthermore, it is assumed that the reference voltage $\Bar{v}$ is known by all inverters, and that $v'_0 = \Bar{v}$.

Now consider the inverter dynamics. A quadratic droop controller can be used as a control law for a PV resource in a smart distribution grid. The quadratic droop controller was first proposed by \cite{dorfler:quad_droop}, and later modified and used in a smart distribution grid by \cite{Chong:initial_droop}. However, while a slope-restriced droop controller was considered in \cite{Chong:initial_droop}, we assume a simplified quadratic droop controller with pure integral control dynamics
\begin{equation}\label{eq:inverter_dynamics}
    \Dot{q}_{g,i} = -K_i(v_i^2 - \Bar{v}^2),
\end{equation}
where $K_i>0$ denotes the inverter gain. Normally, the term proportional to the generated reactive power $-\dfrac{1}{\tau}q_{g,i}$, where $\tau$ is the time-constants of the inverters, is also included in \eqref{eq:inverter_dynamics}. However, this term has been neglected here for simplicity; in other words, it is assumed that the inverters have large time constants.

Next, combine the equations in \eqref{eq:base_eqns} and \eqref{eq:inverter_dynamics} into a state space model. By introducing the output variable $y_i := v_i^2-\Bar{v}^2, i\in\{1,2,\dots,N\}$, and $y$ as the column vector of $y_i$, the system can be rewritten as 
\begin{equation}\label{eq:ssmodel_doubleattack}
    \left\{
    \begin{aligned}
        \Dot{q}_{g} &= Ky,\\
        y &= R' \rho-X'q_c + X'q_g+Ra_\rho+Xa_q,
    \end{aligned}
    \right.
\end{equation}
where $\rho, q_c, q_g, a_\rho, a_q$ denote the column vectors that collect all their respective scalar quantities and $K ~:=~ \text{diag} (-K_1,\dots,-K_N)$. Define $X$, $X'$, $R$ and $R'$ as  
\begin{equation}
X := \begin{bmatrix}
        2X_0 & 2X_0 & \dots & 2X_0\\
        2X_0 & 2(X_0+X_1) & \dots & 2(X_0+X_1)\\
        \vdots & \vdots & \ddots & \vdots \\
        2X_0 & 2(X_0+X_1) & \dots & 2\sum\limits_{i=0}^{N-1} X_i
    \end{bmatrix}
\end{equation}    
with $X' := X+2\;\text{diag}(X'_0,\dots,X'_{N-1})$, and $R$ and $R'$ are defined in a similar manner. Let the attack vectors $a_\rho$ and $a_q$ be the inputs and $q_g$ be the state vector.

\subsection{Assumptions on the Model}\label{subsec:assumptions}
We now introduce some assumptions.

\newtheorem{assumptionImpedance}{Assumption}
\noindent
\begin{assumptionImpedance} \label{th:assumption_impedance}
\hfill 
\begin{enumerate}[i)]

    \item All impedances in the grid satisfy $\frac{R_i}{X_i} = \frac{R'_i}{X'_i} = m$, where $m$ is a constant.

    \item $X'_i = R'_i = 0$, which implies $X' = X, R' = R$.
\end{enumerate}
\end{assumptionImpedance}

A few remarks on Assumption~\ref{th:assumption_impedance} are in order. In a real-world setting, i) states that the same type of cable is assumed to be used in the entire system. In a practical scenario, one would indeed expect that all consumers in a neighborhood are connected at the same time, and it would be practical to use a single type of cable for the entire construction project; hence the assumption would likely hold. Moreover, the assumption approximately holds in the benchmark European low voltage distribution grid in \cite{strunz:benchmark}. Additionally, the assumption implies $R = mX$. Regarding ii), the impedance $Z'_i$ would often represent the impedance of the cable from the edge of the property to the building. In that case, it is reasonable to assume that this cable would be short in comparison to the cables in the rest of the grid, and hence that the impedance from that cable is negligible. 

By only considering deviations from the system's equilibrium, the terms involving $\rho$ and $q_c$ can be disregarded because they are constant. Moreover, applying Assumption~\ref{th:assumption_impedance} to \eqref{eq:ssmodel_doubleattack} yields the simplified state space model 
\begin{equation}\label{eq:ssmodel_final}
    \left\{
    \begin{aligned}
        \Dot{q}_g &= KX q_g + KXa,\\
        y &= X q_g + Xa,
    \end{aligned}
    \right.
\end{equation}
where $a = m a_\rho + a_q$.

\subsection{Attack Scenario}\label{subsec:attack_scenario}
The attack scenario considered is a power injection attack, where an attacker could target any of the $N$ consumer connection points and inject active power $a_{\rho,i}(t)$ and reactive power $a_{q,i}(t)$. It is assumed that the magnitude of the attack is bounded for all $t$ by $|a_{\rho,i}(t) + j a_{q,i}(t)| \le C$, where $C$ is a constant. Moreover, it is assumed that the attack can only be non-zero for $t \in [0,T]$ for some time $T$, and that it is $0$ for all other times. 


There are many ways to implement this attack scenario. All power generation units have a rating which can serve as a maximum bound $C$. Moreover, both PV electronics and electric generators, such as synchronous motors, can control the ratio of generated active and reactive power. Hence, all that is needed to implement a power injection attack is to connect a PV resource or a motor to a power grid.

\subsection{Problem Formulation}\label{subsec:problem_formulation}
In order to analyze the worst-case attack scenario, we consider the following formal problem formulation. The worst attack against a \textbf{line grid} under the attack scenario characterized in Section~\ref{subsec:attack_scenario} is given by the optimization problem
\begin{equation} \label{eq:problem_formulation}
    \begin{array}{ll}
        \argmax\limits_{a,i} & \|y_i\|_{\infty}\\
        \subjectto & \sum\limits_{i=1}^{N} |a_{\rho,i}(t) + j a_{q,i}(t)| \le C,\\
         & a_j(t)=0 , j \neq i,\\
         & t \in [0,T]
    \end{array}
\end{equation}
where $\| y_i \|_\infty$ denotes the $L_\infty$-norm of $y_i$ on the interval $t \in [0,T]$. 
The optimization problem describes a situation where the attacker wants to cause maximal voltage deviation with an attack at one node in the finite time interval $t\in[0,T]$, while constrained by a maximum attack magnitude $C$.

By solving the optimization problem \eqref{eq:problem_formulation}, the following questions can be considered.
\begin{enumerate}[i)]
    \item What is the worst-case attack profile $a(t), t\in [0,T]$?
    \item What is the most vulnerable point in a line grid?
    \item What is the most vulnerable point in a radial grid?
\end{enumerate}
Question i) is answered in Section~\ref{subsec:attack_signal}, question ii) is addressed in Section~\ref{subsec:line_placement}, and finally, question iii) is answered in Section~\ref{subsec:general_placement}. 
Numerical examples are presented in Section~\ref{subsec:verification}. Together, the attack analysis and numerical examples inform a discussion on defense strategies for the grid operator, which is addressed in Section~\ref{subsec:defense}.

\section{Attack Analysis} \label{sec:attack}
This section deals with the three main points of this paper: the worst attack signal, the worst attack location in a line grid, and the worst attack location in a radial grid. 

\subsection{Definiteness of Matrices} \label{subsec:matrix_propositions}
We represent $y(t)$ in convolution form as $y(t) = (g*a)(t)$, where $g(t) := Xe^{KXt}KX+\delta(t)X$ is the system's impulse response, $a(t)$ is the attack signal, and $\delta(t)$ denotes Dirac's delta function. We are primarily interested in $|y_i(T)|$, for some final attack time $T$, which is given by 
\begin{multline} \label{eq:convolution_integral}
    |y_i(T)| = |e_i^{\sf T} y(T)|\\ 
    = \left|\int_0^{T} e_i^{\sf T} Xe^{KX ({T}-\tau)}KX a(\tau) + e_i^{\sf T} \delta(T-\tau)Xa(\tau) d\tau \right|,
\end{multline}
where the second term only depends on time $\tau = T$, and not previous times $\tau \in[0,T)$. Recall that $a_j(\tau) = 0$ for all times if $j \ne i$; in other words that $a(\tau) = e_i a_i(\tau)$.

To draw conclusions about $|y_i(T)|$, we note some properties of the constituent matrices. Since $K$ is a diagonal matrix with negative elements, it is negative definite. We can note similar properties in the matrices $X$ and $R$.

\newtheorem{proposition_definiteness}[assumptionImpedance]{Proposition}
\begin{proposition_definiteness} \label{th:proposition_definiteness}
$X$ and $R$ are positive definite.
\end{proposition_definiteness}
\begin{proof}
The proposition follows as a special case of Lemma 1 in \cite{prop2_proof}, since a line grid is a special case of a radial grid.
\end{proof}

Using Proposition~\ref{th:proposition_definiteness}, we can draw some conclusions about $g(t)$.

\newtheorem{proposition_system_definiteness}[assumptionImpedance]{Proposition}
\begin{proposition_system_definiteness}\label{th:proposition_system_definiteness}
The function $g(t)$ is negative definite for any $t>0$.
\end{proposition_system_definiteness}


\begin{proof}
Begin by introducing a change of basis into \eqref{eq:ssmodel_final} with $z := L^{-1}q_g \iff q_g = L z$, where $L^{-1} :=~ \text{diag}(K_1^{-1/2},\dots,K_N^{-1/2})$.  Now, $g(t)$ can be rewritten as $g(t) = -XL e^{-LXL t} L X$. We want to show that $v^{\sf T} g(t) v = -w^{\sf T} e^{-LXL t} w < 0$, with $w := L X v$. Hence we only need to show that $e^{-LXL t} \succ 0$. By definition, $L$ is symmetric and positive definite, hence $LXL \succ 0$. Moreover, it is possible to diagonalize the exponential factor as $U^{\sf T} e^{-\Lambda t} U$, where $U$ is an orthogonal matrix and $\Lambda$ is a diagonal matrix. We now see that $e^{-\Lambda t} \succ 0$, which since $t>0$ implies $e^{-LXL t} \succ 0$. 
\end{proof}

\subsection{Worst Attack Signal} \label{subsec:attack_signal}
In this subsection, we investigate the worst-case attack signal by considering $|y_i(T)|$ in convolution form, as given in \eqref{eq:convolution_integral}.

The attacker is interested in finding: i) the worst-case attack profile $a(t)$, and ii) the $i$ that gives the maximum voltage deviation. We begin by considering how to choose $a_\rho(t)$ and $a_q(t)$ such that $a(t)$ is maximized.

\newtheorem{lemma_realimag}[assumptionImpedance]{Lemma}
\begin{lemma_realimag} \label{th:lemma_realimag}
Given Assumption~\ref{th:assumption_impedance} and the assumption that the attackers resources are bounded by $|a_{\rho}(t)~ +~ a_{q}(t)|~\le~C$, the maximum attack signal $a(t) = m a_\rho(t) + a_q(t)$ for any $t\in [0,T]$ is time-independent and given by 
\begin{equation}
    \left\{
\begin{aligned}
    a_\rho(t) &= C \cos\left( \arctan \dfrac{1}{m}\right),\\
    a_q(t) &= C \sin \left( \arctan \dfrac{1}{m}\right).
\end{aligned}\right.
\end{equation}
\end{lemma_realimag}



\begin{proof}
Consider the solution to the optimization problem 
\begin{equation}\label{eq:opt_realimag}
    \begin{array}{ll}
        \argmax\limits_{a_\rho(t), a_q(t)} & m a_\rho(t) + a_q(t),  \\
        \subjectto & \sqrt{{a_\rho(t)^2+a_q(t)^2}}{\le C,}
    \end{array}
\end{equation}
for a given $t$. It is clear that \eqref{eq:opt_realimag} is maximized when $\sqrt{{a_\rho(t)^2+a_q(t)^2}}{= C}$. Hence, \eqref{eq:opt_realimag} can be reduced to a single variable optimization problem by introducing polar coordinates with $a_\rho(t) = C \cos \varphi$ and $a_q(t) = C \sin \varphi$, where equality in the constraint holds for all real $\varphi$. The solution to this single variable optimization problem is given by $\varphi = \arctan \frac{1}{m}$, which yields the desired expression.
\end{proof}

With the knowledge of which $a_\rho(t)$ and $a_q(t)$ that maximize $a(t)$ for any $t\in [0,T]$, we are now ready to answer the first question on the worst-case attack profile $a(t)$.

\newtheorem{theorem_worstA}[assumptionImpedance]{Theorem}
\begin{theorem_worstA} \label{th:worstA}
Assume that $|a(t)| \le C_0$. Then, one worst-case finite-time attack signal during $t\in[0,T]$ is given by 
\begin{equation}\label{eq:worst_a}
a(t) = \left\{
\begin{aligned}
    C_0 & , t\in [0,T),\\
    -C_0 & , t=T.
\end{aligned}
\right.
\end{equation}
\end{theorem_worstA}

\begin{proof}
We begin by noticing that $|a(t)|$ is indeed bounded for some $C_0$; this follows by substituting $a_\rho(t)$ and $a_q(t)$ from Lemma~\ref{th:lemma_realimag} into $a(t) = m a_\rho(t) + a_q(t)$.

Now, consider \eqref{eq:convolution_integral}. Notice that its maximum is found if both right hand side terms have the same sign. Since $X$ only has positive elements, the second term has the sign of $a(T)$. From Proposition~\ref{th:proposition_system_definiteness}, it follows the first term has the sign of $-a(t)$. Hence, in order to maximize their sum, $\sign(a(t)) = -\sign(a(T))$ for any $t\in[0,T)$. This implies an attack that switches its sign at the final attack stage.

Finally, note that in order to maximize $|y_i(t)|$, we require $|a(t)| = C_0$ at all times $t\in[0,T]$. Since the sign of $a(t)$ does not impact the magnitude of the voltage deviation, we can assume $a(t) > 0$, and remember that $-a(t)$ gives a voltage deviation with identical magnitude. These arguments combined imply the desired attack signal.
\end{proof}


\begin{figure}
    \centering
    \includegraphics[width = \linewidth, clip, trim=0cm 0cm 1.7cm 0cm]{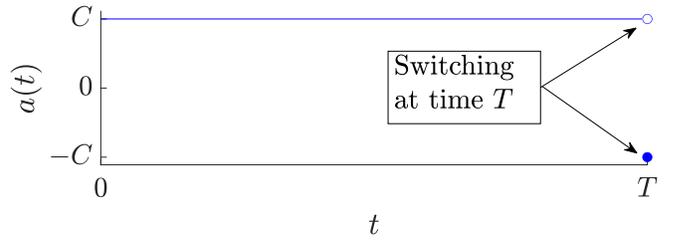}
    \caption{A worst-case attack signal for a finite-time attack during $t\in[0,T]$. The figure shows one of two worst-case attack profiles $a(t)$, the other possibility is the negated $-a(t)$.}
    \label{fig:worstAttack}
\end{figure}

The intuition behind this result is that the system is monotonically driven away from the original equilibrium by a constant maximum amplitude attack at times $t\in[0,T)$ due to Proposition \ref{th:proposition_system_definiteness}. Due to this monotonicity, flipping the attack sign at $t=T$ becomes the worst-case attack. The worst-case attack $a(t)$ from \eqref{eq:worst_a} is visualized in Fig.~\ref{fig:worstAttack}. Because $-a(t)$ gives a voltage deviation of identical magnitude, the negation of \eqref{eq:worst_a} is an equally severe attack.

Now, we have reached our first main conclusion: The worst-case attack signal is given by Theorem~\ref{th:worstA}.

\subsection{Worst Attack Location in a Line Grid} \label{subsec:line_placement}
We are now ready to answer the second question regarding the worst attack location in a line grid.

\newtheorem{theorem_placement_standard}[assumptionImpedance]{Theorem}
\begin{theorem_placement_standard} \label{th:theorem_placement_standard}
The worst attack location in a line grid at the first and final attack stages $t\in\{0,T\}$ as $T \rightarrow \infty$ is at node~$N$.
\end{theorem_placement_standard}


\begin{proof}
Consider the voltage deviation at times $t=\{0,T\}$ as $T \rightarrow \infty$ and assume the attack is the worst case attack signal from \eqref{eq:worst_a} in Theorem~\ref{th:worstA}. Evaluate the convolution integral in \eqref{eq:convolution_integral} at time $t$ 
\begin{equation}
    |y_i(t)| = C_0 e_i^{\sf T} X (-KX)^{-1}\left(I_N-e^{KXt}\right)KX e_i + C_0 e_i^{\sf T} X e_i,
\end{equation}
where the subscript $i$ that maximizes $y_i$ is sought. By definition, the largest element of the matrix $X$ is $[X]_{N,N}$, which implies that the second term is maximized by $i=N$. 

For the first term, consider both $t = 0$ and $t=T$ as $T \rightarrow \infty$. If $t=0$, then $I_N-e^{KXt} = 0$, which implies $i=N$ maximizes $|y_i|$. In the case of $t=T$ as $T \rightarrow \infty$, then $I_N-e^{KXt} \rightarrow I_N$, which yields $|y_i| = 2 C e_i^{\sf T} X e_i$, and $i=N$ maximizes $|y_i|$ again. Thus, it has been shown that $i=N$ produces the largest voltage deviation for the first and last attack stages.
\end{proof} 

This result is intuitively reasonable. If the attack is targeted against the first node, then the injected power flows through $Z_0$, which causes the voltage to deviate from the previous equilibrium. The further away from the substation, the more severe the attack will be, since each additional impedance further increases the deviation from the equilibrium. 

\newtheorem{remark_times}[assumptionImpedance]{Remark}
\begin{remark_times}
It is difficult to say anything in general about the worst attack location for times $t\in(0,T)$ because this depends on impedances in the grid, as well as the inverter dynamics given by $K_i$. This will be illustrated later in Fig.~\ref{fig:examples}, where an attack at node $N$ produces the largest deviation at times $t\in\{0,T\}$, but not at all times in between. 
\end{remark_times}

\newtheorem{remark_placement}[assumptionImpedance]{Remark}
\begin{remark_placement} \label{th:remark_placement}
For a line grid, the voltage deviation of the worst case attack from Theorem~\ref{th:worstA} is proportional to the \textit{electrical distance} to that node; $|y_N(t)| \propto \sum\limits_{i=0}^{N-1} X_i$ for $t\in\{0,T\}$. The constants of proportionality are $2C$ when $t=0$, and $4C$ when $t = T$ as $T \rightarrow \infty$.

Now, we have reached our second main conclusion: The node furthest from the substation in a line grid, node $N$, is the most vulnerable node.
\end{remark_placement}

\subsection{Worst Attack Location in a General Radial Grid} \label{subsec:general_placement}
In this section, we extend the result of Theorem~\ref{th:theorem_placement_standard} to a radial grid. We will prove the intuitively reasonable result that the most vulnerable node in a radial grid is the node with the largest electrical distance to the substation. While the extension might seem trivial, recall that the grid model in \eqref{eq:base_eqns}, and hence all theorems thus far, only apply to line grids, and not to radial grids. However, note that the grid between two branching points in a radial grid can be seen as a line grid. This insight, coupled with the conclusion from Remark~\ref{th:remark_placement}, enables us to extend the result of Theorem~\ref{th:theorem_placement_standard} to a general radial grid topology.

Now, we are ready to answer the third question on the most vulnerable node in a radial grid.

\newtheorem{theorem_placement_general}[assumptionImpedance]{Theorem}
\begin{theorem_placement_general} \label{th:theorem_placement_general}
Consider a radial grid topology and assume that the attacker uses the worst-case attack signal in \eqref{eq:worst_a}. Then the most vulnerable node is the node with the largest impedance to the substation.
\end{theorem_placement_general}


\begin{proof}
Recall that a radial grid can be seen as a collection of line grids and branching points. Here, we use the term ``node'' to denote a line grid within the radial grid. Begin by numbering all nodes in the tree, starting with the root node (which includes the radial grid substation) as node $i=0$. Then, for all nodes, introduce the triple $T_i = (T_{i,1}, T_{i,2}, T_{i,3}) := (i,\Sigma_i, i_\text{max})$, where $i_\text{max}$ denotes the most vulnerable node in the sub-tree to which node $i$ is the root node, and $\Sigma_i$ denotes the vulnerability of the node, which is specified below. Notice that at the leaf nodes, the triple is $T_N = (N,\Sigma_N, N)$, with $\Sigma_N$ as the maximum electrical distance within the node due to Remark \ref{th:remark_placement}. We will prove this theorem using a recursive approach starting from the leaf nodes, since $T_N$ is known for all leaf nodes.

Now let node $i$ be any node in the tree except for the root node, and let node $j$ be its parent. Moreover, assume that $T_i$ is known. Notice that node $j$ can have $k$ number of children, one of which is node $i$. If $k=1$ for all nodes in the grid, all nodes can be combined into a single line grid, which results in a trivial extension of Theorem~\ref{th:theorem_placement_general}. On the other hand, consider the case that $k\ne 1$ for some nodes.

If $k\ne 1$, then first consider how to choose the most vulnerable of the $k$ branches. Let the set of child nodes to node $j$ be given by $S_{jk} := \{T_{k}\}$. Then, the triple associated to the most vulnerable node in $S_{jk}$ is given by $T_\text{max} := T_m$, where $m$ is given by $m = \argmax\limits_{T_m\in S_{jk}} T_{m,2}$. Due to Remark~\ref{th:remark_placement}, the maximum voltage deviation of the most vulnerable branch is proportional to $T_{m,2}$. Hence, the triple for node $j$ is given by $T_j = (j,\Sigma_j+T_{\text{max},2},T_{\text{max},3})$. 

By applying this argument recursively starting with the leaf nodes, the triple at the substation will be $T_0 = (0,\Sigma_0, i_{\text{max,grid}})$, and the most vulnerable node in the grid is then known through $i_{\text{max,grid}}$. Moreover, it will be the node with the largest electrical distance to the substation due to the back propagation of $\Sigma_0$.
\end{proof}

Now, we have reached the final main conclusion: In a grid with radial topology, the node with the largest electrical distance to the substation is the most vulnerable.

\begin{table}
    \centering
    \caption{Benchmark model parameters for a line grid with $N=5$.}
    \label{tab:model_parameters}
    \begin{tabular}{cccccc}
        i & 1 & 2 & 3 & 4 & 5  \\
        \hline \hline \\[-5pt]
        $R_{i-1}$ [$\Omega$] & 0.00343 & 0.00172 & 0.00343 & 0.00515 & 0.00172 \\ 
        $X_{i-1}$ [$\Omega$] & 0.04711 & 0.02356 & 0.04711 & 0.07067 & 0.02356 \\ 
        $R'_{i-1}$ [$\Omega$] & 0.00147 & 0.00662 & 0.00147 & 0.00147 & 0.00147 \\ 
        $X'_{i-1}$ [$\Omega$] & 0.02157 & 0.09707 & 0.02157 & 0.02157 & 0.02157 \\ 
        $K_i$ [s] & 1 & 1 & 1 & 1 & 1\\
        $\rho_{i}$ [W] & 1205 & -60 & 1440 & 2205 & 280 \\
        $q_{c,i}$ [VAr] & 300 & 960 & 480 & 600 & 400 \\[3pt]
        \hline
    \end{tabular}
\end{table}

\section{Simulations and discussion} \label{sec:simulations}
In this section, we provide numerical examples and discuss defense strategies for the grid operator; in Section~\ref{subsec:verification} we verify our theoretical results, and defense strategies are discussed in Section~\ref{subsec:defense}.

\subsection{Verification of Theoretical Results}\label{subsec:verification}
\begin{figure}
    \centering
    \includegraphics[width=\linewidth, clip, trim=0cm 0.4cm 0cm 1.3cm]{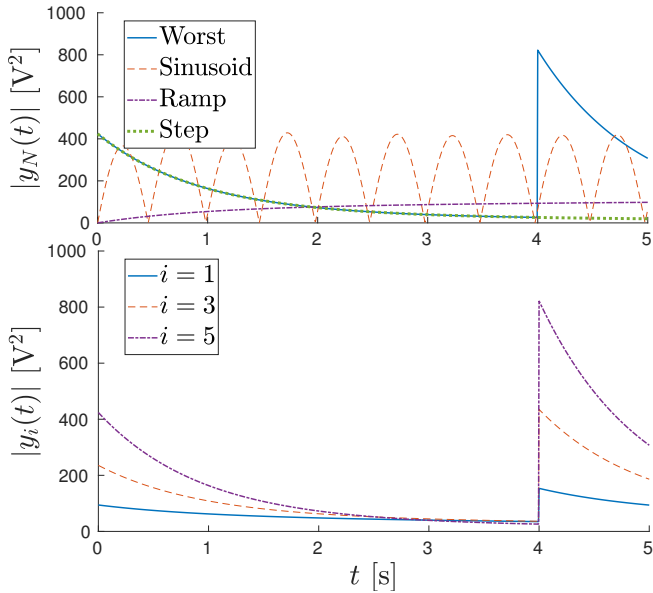}
    \caption{Comparison of different attack scenarios. Top compares different attack signals, and bottom compares different attack locations.}
    \label{fig:examples}
\end{figure}

In this subsection, we provide numerical verification of our results on a benchmark residential European low voltage distribution network from \cite{strunz:benchmark}, which was also used in \cite{Chong:initial_droop}. The grid is a line grid with $N=5$, where grid parameters from \cite{Chong:initial_droop} have been used, and the simplified droop controller is assumed to have $K_i = 1$ for all $i$, see Tab.~\ref{tab:model_parameters}. Since $R_i/X_i = R'_i/X'_i \approx m$ for all $i$, the system \eqref{eq:ssmodel_final} is considered, and it is assumed that $a_\rho$ and $a_q$ are given by Lemma~\ref{th:lemma_realimag}. Additionally, $C=1000$ is assumed to be the attacker's maximum attack strength. With these model parameters, two comparisons have been made: a comparison between different attack signals, and between different attack locations, see Fig.~\ref{fig:examples}.

The impact of different attack signals have been compared in Fig.~\ref{fig:examples}a: the worst case attack signal from \eqref{eq:worst_a}, a sinusoid signal, a ramp signal, and a step function have been compared. The maximum quadratic voltage deviation $y(t)$ for node $N=5$ has been considered, and all compared signals have the same maximum amplitude $C=1000$. The derived worst case attack signal from \eqref{eq:worst_a} is clearly the worst of the considered attack signals; at $T=$ 4 \rm{s}, switching occurs and the voltage deviation increases. On the other hand, if the switching occurs quickly, for small $T$, then the system will still be close to its original equilibrium. Therefore, a fast switching will result in a voltage deviation of a similar magnitude as that of the initial attack at $t=0$.

The impact of attacks at different nodes has been compared in Fig.~\ref{fig:examples}b. As expected, the worst case attack location is the $N$-th node. Notice that between times $0$ and $T$, it is difficult to say anything about the node with the largest voltage deviation; right before $t=4$, node $i=3$ gives the largest voltage deviation. This is the reason why only times $t=\{0,T\}$ are considered in Theorem~\ref{th:theorem_placement_standard}; depending on impedances in the grid, and time parameters $K_i$, different nodes will have different voltage deviations for times $t\in~ (0,T)$.

\subsection{Defense Strategies} \label{subsec:defense}
In this subsection we provide a brief discussion on possible defense strategies for the grid operator.

\begin{figure}
    \centering
    \includegraphics[width = 0.8\linewidth, clip, trim=0.4cm 1cm 1.7cm 0.9cm] {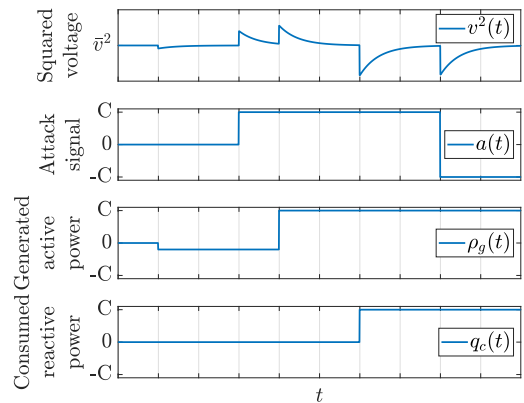}
    \caption{Shows that it is difficult to distinguish normal behavior from an attack; normal behavior result in changes in $\rho$ and $q_c$, whose impact on the voltage deviation is equivalent to an attack.}
    \label{fig:unfeasible}
\end{figure}

Firstly, consider the impact of $\rho$ and $q_c$ on voltage deviation, see Fig.~\ref{fig:unfeasible}. The line grid considered is the simple scenario where $N=1$. All other parameters are chosen to illustrate a fundamental problem for the grid operator: in general, without additional assumptions on available measurement data or PV specifications, it is difficult to distinguish the worst case attack from normal grid operation. This problem would be common in a real-world setting, since many household items draw near-constant power once turned on (e.g. microwaves, stoves or electric vehicles), and hence their power consumptions are on the form of step functions. While assuming an upper bound on $\rho$ and $q_c$ might be useful to design a detector, such assumptions might yield additional issues in practice: electric vehicles require large amounts of power when charging; therefore, an upper bound on power consumption might be too large to be practically useful in a detection algorithm. 

Due to the difficulty of attack detection, a detector will not be derived in this paper. Instead, we focus on grid reconfiguration algorithms and grid design. 

We begin by considering grid design. Based on the analysis in Theorem~\ref{th:theorem_placement_general}, power injection attack resilient radial distribution grids should minimize the electrical distance between the substation and consumers. In addition, in order to cope with attacks, the grid should be designed with flexibility so that grid reconfiguration algorithms are possible and effective. 

Grid reconfiguration is a difficult problem which has been studied in e.g. \cite{bernadon:2012_reconfiguration, lopez:2015_reconfiguration, trpovski:2017_reconfiguration}. A common approach is to consider an objective function that minimize resistive losses in the grid. In order to minimize power injection attack vulnerability in a radial grid, the maximum electrical distance needs to be minimized according to Theorem~\ref{th:theorem_placement_general}. Two intuitive approaches to incorporate the conclusion from Theorem~\ref{th:theorem_placement_general} into existing research are: either adding a second objective function that minimizes electrical distance, or to minimize the maximum electrical distance in the grid instead of minimizing resistive losses. At a glance, the latter approach could be approximately equivalent to the existing objective function, since reducing the electrical distance likely reduces resistive losses in the grid as well.

\section{Conclusion} \label{sec:conclusion}
In this paper, we have considered the problem of an attack against the physical layer of a smart distribution grid. We reached three main conclusions: the worst-case finite-time attack signal is given by a step function that switches its sign at the final time of the attack, and the worst attack location in both a radial grid and a line grid is at the node with the largest electrical distance to the substation. Moreover, we have argued that detection of the worst case attack is very difficult with access only to voltage levels in the grid. Hence, we conclude that grid design and grid reconfiguration algorithms need to be considered in order to mitigate the severity of power injection attacks. Further research is needed to incorporate attack resilience against power injection attacks into grid reconfiguration algorithms. 

\bibliographystyle{IEEEtran}
\bibliography{martinref}
\end{document}